\documentclass[aps,prb,twocolumn,showpacs,floatfix]{revtex4}
\usepackage{graphicx}
\usepackage{bm}
\usepackage{amsmath,amssymb}

\def\bea{\begin{eqnarray}}
\def\eea{\end{eqnarray}}
\def\beann{\begin{eqnarray*}}
\def\eeann{\end{eqnarray*}}

\def\lrb{\left(}
\def\rrb{\right)}
\def\lcb{\left\{}
\def\rcb{\right\}}

\def\ii{\mathrm{i}}
\def\Im{\mathrm{Im}}

\def\mymathbf{\boldsymbol}

\begin{document}

\title{Hofstadter butterflies of carbon nanotubes:\\Pseudofractality of the magnetoelectronic spectrum}

\author{Norbert Nemec}
\author{Gianaurelio Cuniberti}
\affiliation{Institute for Theoretical Physics,
University of Regensburg,
D-93040 Regensburg, Germany}

\date{August 29, 2006}

\begin{abstract}
The electronic spectrum of a two-dimensional square lattice in a
perpendicular magnetic field has become known as the Hofstadter
butterfly [Hofstadter, Phys. Rev. B \textbf{14}, 2239 (1976)]. We have
calculated quasi-one-dimensional analogs of the Hofstadter butterfly for carbon nanotubes (CNTs).
For the case of single-wall CNTs, it is
straightforward to implement magnetic fields parallel to the tube axis by
means of zone folding in the graphene reciprocal lattice. We have also
studied perpendicular magnetic fields which, in contrast to the parallel
case, lead to a much richer, pseudofractal spectrum. Moreover, we have
investigated magnetic fields piercing double-wall CNTs and found strong
signatures of interwall interaction in the resulting Hofstadter butterfly spectrum, which can be
understood with the help of a minimal model.
Ubiquitous to all perpendicular magnetic field spectra is the presence
cusp catastrophes at specific values of energy and magnetic field.
Resolving the density of states along the tube circumference allows recognition of the
snake states already predicted for nonuniform magnetic fields in the two-dimensional electron gas.
An analytic model of the magnetic spectrum of electrons on a cylindrical surface is used to explain some
of the results.
\end{abstract}

\pacs{
73.63.Fg,
73.22.-f,
73.43.-f,
73.43.Qt
}

\maketitle

\section{Introduction}

The availability of new materials for nanoelectronic research allows for a
detailed test of the emergence of the quantum physical nature of electrons, via
transport or optical measurements. Carbon
nanotubes{\cite{iijima-hmogc1991,oberlin-fgoctbd1976,dresselhaus-sofacn1996,saito-ppocn1998,reich-cnbcapp2004,thune-qticn2005}}
(CNTs)
are an example of a very peculiar electronic material, due to the
extreme confinement of electrons on their $\pi$-conjugated ``walls''.
In these systems, many mesoscopic phenomena such as
single-electron charging,\cite{postma-cnstart2001}
and conductance quantization,\cite{frank-cnqr1998} as well as effects typical
for semiconductor physics like $s$-like excitons,\cite{wang-toricnafe2005}
can be observed already at room temperature.

Since the prediction of band structure effects of carbon nanotubes in parallel
external fields by Ajiki and Ando in 1993,\cite{ajiki-esocn1993}
it took only a few years until clear
hallmarks of a single quantum flux being tethered within a tube section were
found experimentally in optical\cite{zaric-osotapiscn2004} and
transport\cite{bachtold-aoicn1998, minot-doeommicn2004} measurements.
For magnetic fields perpendicular to the CNT axis, theoretical predictions
were made shortly after, first using a perturbative approach around the
Fermi energy,\cite{ajiki-mpocn1993} and later also
using a tight-binding model.\cite{saito-mebocn1994, ajiki-ebocnimf1995}
Only recently, a first experimentally accessible effect of perpendicular
magnetic fields---anomalous magnetoconductance---was
predicted\cite{roche-mocnfmtmf2001} and observed.\cite{fedorov-gmpicn2005}
A very similar effect for strong electric fields has also been found by
numerical studies \cite{son-esimcn2005} and has yet to be confirmed
experimentally.
The use of magnetic fields to further investigate the interplay between
elastic mean free path, phase coherent length, and electron-electron
interaction was also successfully adopted.\cite{roche-asfaclicn2000,stojetz-eobsoqiimcn2005,stojetz-cbmfdbsacbimcn2006}

From the purely theoretical perspective, carbon nanotubes in strong
perpendicular magnetic fields represent a very interesting case of study. Closely related
to graphene, their energy spectrum shows strong similarities with that of the
two-dimensional (2D) honeycomb lattice,\cite{rammal-llsobeiahl1985,kreft-moth1996, pedersen-ttofrig2003} which again
forms a variation of the fractal butterflylike pattern discovered by
Hofstadter\cite{hofstadter-elawfobeiraimf1976} in 1976 and studied intensely since that time from various points
of view.\cite{albrecht-eohfesitqhc2001, osadchy-hbaqpd2001, iye-hbiammfswnwmd2004,brning-hsdftbeitd2004, analytis-llmoathbifs2004, zhou-eoldothb2005}
Yet the quasi-1D nature and the curvature of CNTs set their energy spectra clearly apart from the
fractal and perfectly periodic images obtained in 2D lattices.

\begin{figure}[b]
\includegraphics[width=\columnwidth]{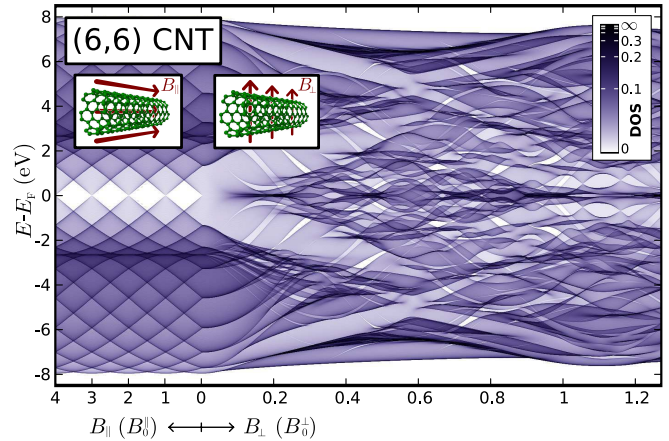}
\caption{\label{fig:firstplot}(Color online) Density of states of a (6,6) CNT in dependence
on an external magnetic field parallel (left) or perpendicular (right) to the
tube axis. For every value of the magnetic field, the DOS is unity normalized over energy.
The units $B_0^{\parallel}=\Phi_0/r^2 \pi$ and $B_0^{\perp}=\Phi_0/A_{\text{plaquette}}$ (see text) are scaled
such that the physical field scale is the same for both segments of the plot.
}
\end{figure}

In this paper, we will describe a method of computing and visualizing the
spectrum of carbon nanotubes (for a prototypical example see Fig.~\ref{fig:firstplot}).
This method will be demonstrated on a number of
single- and double-wall CNTs (SWCNTs and DWCNTs) of different chirality and diameter.
The study of the local distribution of the spectral density will shed some light on the
relation between the spectrum of a planar sheet of graphene and that of a
CNT, strongly affected by curvature and finite size.
A closer look at the spectrum will reveal the presence of cusp catastrophes,
which are closely related to the quenching of the Bloch state velocity,
induced by a magnetic field.

\begin{figure}[tb]
\includegraphics[width=\columnwidth]{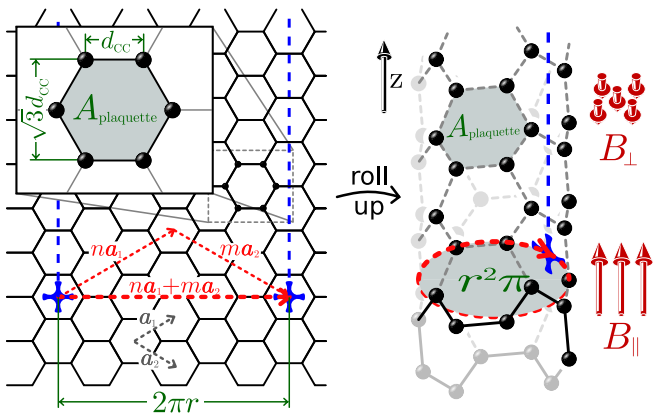}
\caption{\label{fig:rollup}(Color online) 
The structure of a CNT [here, a (3,3)-CNT]: the hexagonal lattice of a
graphene sheet is rolled up in such a way that the chiral vector
$(n,m)$ becomes the circumference of the resulting cylinder. Magnetic
fields parallel to the tube axis pierce the tube cross section $r^2\pi$,
while perpendicular magnetic fields pierce the wall made up from
hexagonal plaquettes.
}
\end{figure}

For magnetic fields parallel to the tube axis, the natural unit is that of one
flux quantum per tube cross section $r^2 \pi$ (see Fig.~\ref{fig:rollup}). For a
general $(n,m)$~CNT the tube radius can be obtained with simple geometrical arguments
$ 2 \pi r = \sqrt{ 3 m^2 + 3 n^2 + 3 m n }\, d_\mathrm{CC}$,
where $d_\mathrm{CC}=1.42~\text{\AA}$ denotes the carbon-carbon distance.
This immediately gives the parallel magnetic field $B_0^{\parallel}$
needed to pierce one flux quantum
$\Phi_0 = h/e$ through an $(n,m)$ CNT.
For perpendicular magnetic fields, the scale is ruled by the field necessary to enclose a flux through
a single benzene ring, the plaquette of graphene and carbon nanotubes of
area $A_{\text{plaquette}} = 3 \sqrt{3/4} ~ d_{\rm CC}^2 \sim 5.24 ~ \text{\AA}{}^2$.
Because of this
extremely small area we obtain $B_0^{\perp} = \Phi_0 / A_{\text{plaquette}} =
79\times10^3 \operatorname{T}$,
which is, of course, out of experimental reach.\cite{herlach-pm1999}
It is straightforward to get the relation between the parallel and perpendicular field scales as
\begin{equation}
B_0^{\parallel} = \frac{\Phi_0}{r^2 \pi} = \frac{2 \sqrt{3} \pi}{m^2 + n^2 + m n} B_0^{\perp}~.
\end{equation}
For a typical SWCNT with ${\sim}1~\operatorname{nm}$ diameter, this gives a value
of $B_0^{\parallel} \approx 5\times10^3 \operatorname{T}$.
It is thus understandable that multiwall CNTs (MWCNTs) present a more interesting object for magnetic field experiments:
For a typical MWCNT with a diameter of 20~nm, as a matter of fact, one can already
observe the first Aharonov-Bohm oscillations accessible at
around 12~T parallel fields.\cite{bachtold-aoicn1998}
As shown in this work, however, even for perpendicular fields low-field signatures 
could be visible within experimentally accessible field ranges if one
takes into account the external shell of a MWCNT.

This paper is organized as follows. We first give definitions, introduce the method of computation and
visualization, and point toward general features observable in quasi-1D systems. In Sec.~III, we then
do a systematic study of SWCNTs, including an analytic model and a detailed view of the range of experimentally
accessible fields. In Sec.~IV, we proceed with an analysis of the effects of the interwall interaction in
DWCNTs on the magnetic spectrum and introduce a minimal model, closing with
a discussion of the results in the last secion.

\section{Definitions, methods, and observables}

\begin{figure*}[t]
\includegraphics[width=\textwidth]{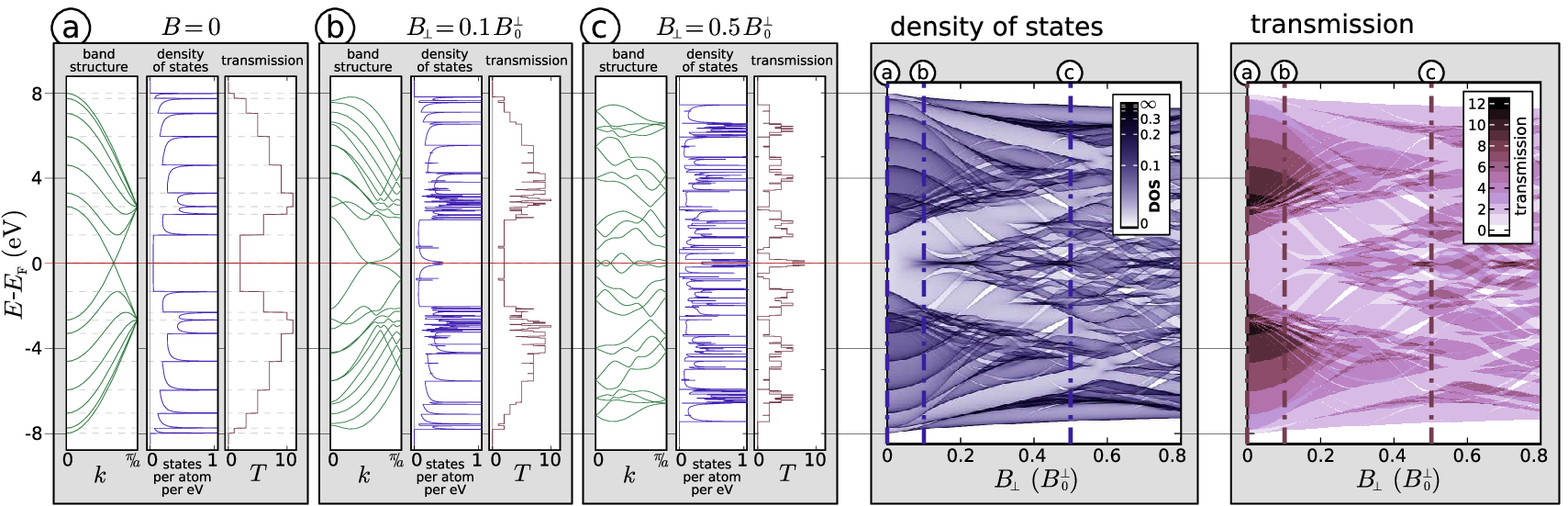}
\caption{\label{fig:explain}(Color online) Scheme to illustrate the physical
meaning of the butterfly plots. An external magnetic field
distorts the band structure of a CNT in an intricate way. For any
fixed magnetic field, the DOS and the transmission can be obtained
directly from the band structure. (a), (b), and (c) are sections of
the two right panels of the DOS and transmission vs $E$ and
$B_{\perp}$.
}
\end{figure*}

\subparagraph*{Lattice electrons in arbitrary external magnetic fields.}

All numerical calculations in this work are based on a tight binding Hamiltonian of the
form
\begin{eqnarray*}
\mathcal{H}(\mymathbf{B}) & = & \sum_i \varepsilon_i^{\phantom{\dag}} c_i^{\dag}
c_i^{\phantom{\dag}} - \sum_{\left < i, j\right >} \gamma^{\phantom{\dag}}_{i j}(\mymathbf{B}) c_i^{\dag}
c_j^{\phantom{\dag}},
\end{eqnarray*}
where the indices denote the atomic orbitals. For the single-orbital
approximation used hereafter, these coincide with the label of the atom so
that $\mathcal{H}$ can be represented by means of the matrix
elements $\mathcal{H}_{\mymathbf{r}_i,\mymathbf{r}_j}$ between $\pi$ orbitals
centered on the atom $i$ and $j$ at the position $\mymathbf{r}_i$ ($\mymathbf{r}_j$) of the CNT molecular
network.

An external magnetic field is implemented using the Peierls substitution:\cite{peierls-ztddvl1933}
Based on the principle of minimal coupling $\mymathbf{p} \rightarrow \mymathbf{p} - e \mymathbf{A}$, the
effect of a magnetic field $\mymathbf{B} = \operatorname{rot} \mymathbf{A}$ is absorbed in the
translation operator
$\mathcal{T}\left( \mymathbf{R}\right ) = \exp \frac{\ii}{\hbar}\left(\mymathbf{p}-e \mymathbf{A}\right)\cdot\mymathbf{R}$.
In the tight-binding Hamiltonian, this is reflected by representing the hopping matrix elements 
$\gamma_{i j} = \left <\Psi_i \vert \mathcal{H} \vert \Psi_j \right >$
between two $\pi$ orbitals
$\Psi_i$ and $\Psi_j$ localized at sites $\mymathbf{r}_i$ and $\mymathbf{r}_j$ as
\begin{eqnarray}
\gamma_{i j}^{\phantom{0}}( \mymathbf{B} ) & = & \gamma_{i j}^0 \exp \left( \mathrm{i}
\frac{2 \pi}{\Phi_0} \int_{\mymathbf{r}_i}^{\mymathbf{r}_j} \mathrm{d} \mymathbf{r}
\cdot \mymathbf{A}_{\mymathbf{B}} \left( \mymathbf{r} \right) \right)~.\label{peierls}
\end{eqnarray}
The bare hopping at zero magnetic field $\gamma_{ij}^0$ acquires a complex phase expressed as an integral
along the bond direction $\mymathbf{d}_{ij} = \mymathbf{r}_j - \mymathbf{r}_i$.

With the CNTs oriented parallel to the $z$ axis, it is advantageous to choose a gauge
in such a way that $\mymathbf{A}_{\mymathbf{B}}$ is independent of $r_z$. This is provided, e.g., by
\begin{eqnarray}
\mymathbf{A}_{\mymathbf{B}} \left( \mymathbf{r} \right) = \left(
0, r_x B_{\parallel}, r_y B_{\perp}
\right), \label{gauge-field}
\end{eqnarray}
giving a magnetic field $\mymathbf{B}=(B_{\perp},0,B_{\parallel})$
with known components perpendicular and parallel to the tube axis.
Throughout this work fields will be consider either perpendicular
($B_{\parallel}=0$) or parallel ($B_{\perp}=0$) to the tube axis.
Arbitrary angles are of course possible as well, showing the
expected crossover of both regimes.

Having chosen a linear gauge further simplifies the integration in Eq.
(\ref{peierls}) to a product:
\begin{eqnarray}
\gamma_{i j}^{\phantom{0}} & = & \gamma_{i j}^0 \exp \left[ \mathrm{i}
\frac{2 \pi}{\Phi_0} \mymathbf{d}_{ij} \cdot
\mymathbf{A}_{\mymathbf{B}} \left( \frac{\mymathbf{r}_j + \mymathbf{r}_i}{2} \right)
\right]. \label{peierls-gauge}
\end{eqnarray}

In the presence of a perpendicular magnetic field, it is thus necessary to consider
the exact coordinates of the molecular structure at hand rather than---as sufficient
for parallel or vanishing magnetic fields---their simple topological connectivity.

\subparagraph*{Density of states.}

For such an $r_z$-independent gauge field, the Hamiltonian of any quasi-1D
periodic structure like a CNT stays periodic in the presence of a
magnetic field. This allows the use of the Bloch theorem to derive the corresponding
band structure. As can be seen in Fig.~\ref{fig:explain}, the band structure is
in general strongly distorted by an applied magnetic field. The density of
states (DOS) can be determined from the magnetic band structure $E_b (k, \mymathbf{B})$ via
\begin{eqnarray}
\rho_{\operatorname{DOS}} (E, \mymathbf{B}) & = & \frac{a}{2 \pi N_{b}} \sum_{b=1}^{N_{\mathrm{b}}}
\int_{-\pi/a}^{\pi/a} \mathrm{d} k ~
\delta \left( E - E_b (k, \mymathbf{B}) \right) \label{DOS-definition}
\end{eqnarray}
where $b$ is the band index and
$a=3 d_{\rm CC} \sqrt{m^2+n^2+mn}/\operatorname{gcd}\left (3n,n-m\right )$
the length of the unit cell of an
$(n,m)$ CNT. $E_b (k, \mymathbf{B})$ is obtained by direct diagonalization of the 
CNT Hamiltonian via the Bloch ansatz (see Appendix~\ref{app:histogram}).
Since we work in a basis of one orbital per atom, the number of bands $N_\mathrm{b}$ equals the number of
atoms in the unit cell $N=4 \left (n^2+m^2+nm\right )/\operatorname{gcd}\left (3n,n-m\right )$,
growing with the diameter and dependent on the helicity angle (deviation from the armchair or zigzag configuration).
The plot of the DOS directly reflects the distortions of the band structure caused by the magnetic field.

Alternatively, Green-function-based approaches allow one
to resolve the DOS within different atoms in the same unit cell by
introducing the local density of states
\begin{eqnarray}\label{LDOS-definition}
\rho_{\operatorname{LDOS}_{\mymathbf{r}_i}}(E, \mymathbf{B}) =
-\frac{1}{\pi} \Im\ \mathcal{G}_{\mymathbf{r}_i,\mymathbf{r}_i} \lrb E,
\mymathbf{B}\rrb,
\end{eqnarray}
where
$\mathcal{G}_{\mymathbf{r}_i,\mymathbf{r}_i}$
is the space-diagonal
component of the lattice
Green function matrix (see Appendix~\ref{app:greenfcn})
\begin{eqnarray}\label{lattice-GF}
\mathcal{G} \lrb E, \mymathbf{B}\rrb = \left[ E -
\mathcal{H}(\mymathbf{B}) + \mathrm{i}0^+ \right]^{-1}.
\end{eqnarray}
Of course by tracing the LDOS within the different atoms of the same unit cell, one can restore the full
DOS as
$\rho_{\operatorname{DOS}}(E,\mymathbf{B}) = \frac{1}{N}\sum_{\mymathbf{r}_i} \rho_{\operatorname{LDOS}_{\mymathbf{r}_i}}(E,\mymathbf{B})$.

\begin{figure*}[t]
\includegraphics[width=0.8\textwidth]{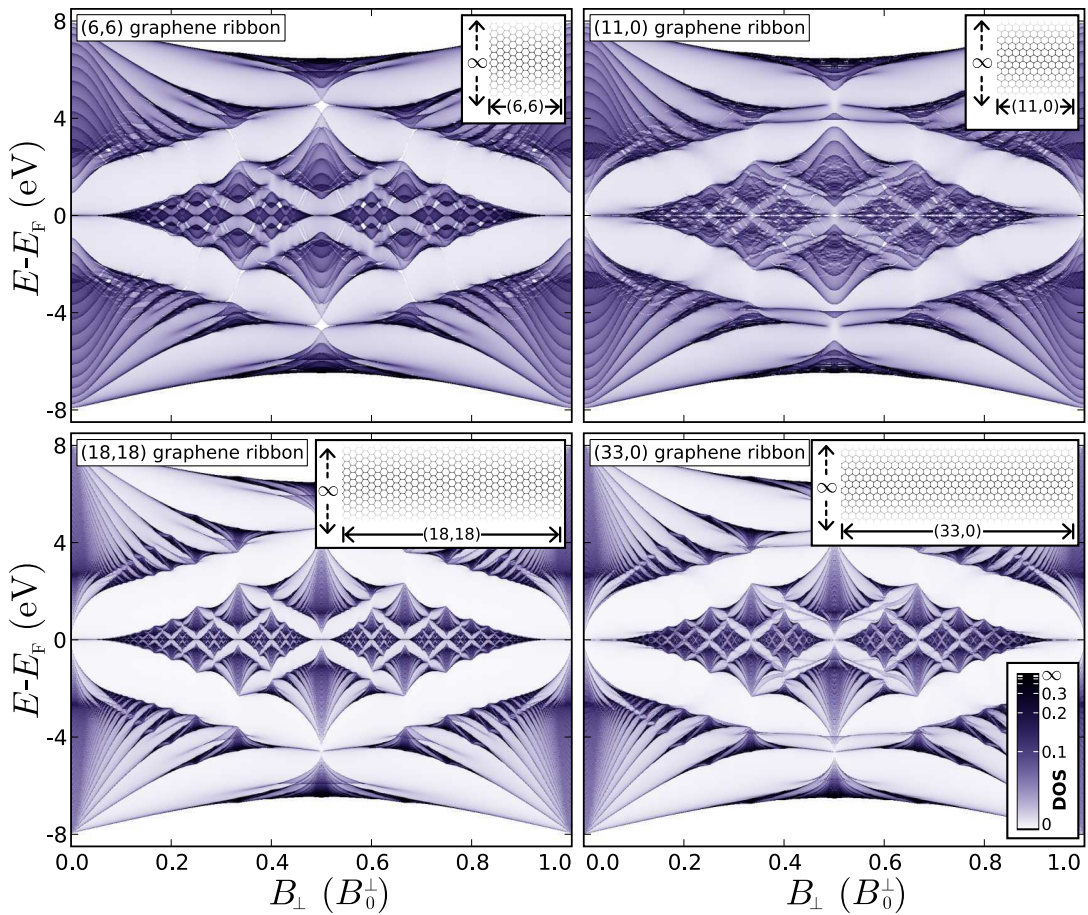}
\caption{\label{fig:grapheneribbon}(Color online) DOS in a graphene ribbons of infinite length
and various widths and internal orientations, pierced perpendicularly by magnetic
fields. Each ribbon can be classified as an unrolled CNT:
The ``chiral'' vectors refer to the SWCNT which,
when unrolled, would result in the corresponding planar ribbon. The
density of states is normalized to the number of atoms per unit
cell to give a comparable visual appearance.
}
\end{figure*}

\subparagraph*{Butterfly plots.}

To capture the continuous evolution of the band structure with
growing magnetic fields, it is very convenient to visualize the DOS
in butterfly plots, as illustrated in Fig.~\ref{fig:explain}.
The resemblance to the well-known Hofstadter butterfly
of 2D lattice electrons\cite{hofstadter-elawfobeiraimf1976} becomes
very clear for CNTs of large diameter (see
Fig.~\ref{fig:diameter} below). A common feature to butterfly plots of
all quasi-1D systems are the pronounced band edges, caused by
van Hove singularities in the DOS.\cite{hove-toositefdoac1953}

In Fig.~\ref{fig:firstplot}, a (6,6) CNT Hofstadter butterfly is
plotted as a reference for further comparisons. For the parallel
field, the behavior is perfectly periodic for integer multiples of
the flux quantum $\Phi_0 = h / e$ penetrating the tube cross section
$r^2 \pi$. Starting as a metallic CNT at $B = 0$, the gap opens and
closes periodically.\cite{ajiki-esocn1993}

For perpendicular fields with their natural scale of one flux quantum per
graphene plaquette, the overall
behavior is not periodic. This can be understood due to the presence
of plaquettes at various angles toward the field, capturing
different, in general incommensurate, fractions of the flux quantum.
However, a number of features from the underlying graphene structure
are still visible at the diameter-independent scale of $B_0^\perp$.

Important to note is the difference in the behavior for small fields: while
the parallel field causes a linear Zeeman split of the states with opposite
angular momentum, small perpendicular fields generally cause quadratic energy
shifts.

All plots are of course symmetric in the magnetic field sign, which is why only half butterfly plots are shown. Different is the case of the $E\rightarrow-E$ symmetry which is related to the particle-hole symmetry. The latter is present in the $\pi$-orbital description of
SWCNTs but is broken by the interwall interaction in DWCNTs.

\subparagraph*{Transport observables in quasi-1D systems.}

As can also be seen in Fig.~\ref{fig:explain}, it is
straightforward to apply the same  scheme not only to
the density of states, but just as well to other properties like the
quantum mechanical transmission $T$ of a quasi-1D system. The latter is the dimensionless
zero-temperature conductance after the Landauer theory of phase-coherent
transport:\cite{datta-qtatt2005} $G=G_{\rm K}T$, where $G_{\rm K} =
2e^2/h$ is the conductance quantum and inverse of the von~Klitzing
resistance. The calculation of the transmission, which involves a
renormalization procedure for the semi-infinite carbon nanotube
leads\cite{cuniberti-trocime2002} by means of the energy-dependent
injection rates $\Gamma_{\mathrm{L}/\mathrm{R}}$ and the Green
function $\tilde{\mathcal{G}}$ projected on a finite nanotube partition, can be cast
into the Fisher and Lee formula\cite{fisher-rbcatm1981}
\begin{equation*}
G=\frac{2e^2}{h} \textrm{Tr} \lcb
\Gamma_{\mathrm{L}}\tilde{\mathcal{G}}\Gamma_{\mathrm{R}}\tilde{\mathcal{G}}^\dagger
\rcb.
\end{equation*}

Still, for a periodic structure---as is the case for the systems at hand---the
quantum mechanical transmission is simply a band-counting algorithm, and as such
contains less information than the band structure itself or the DOS. This is very different
from magnetotransport through finite CNTs: Scattering at the contacts leads to resonant
tunneling, resulting in spectroscopy of the electronic states of the finite
tube.\cite{krompiewski-ieiettmscn2002,krompiewski-gmomcnmttbc2004} This spectrum
may show strong dependence on magnetic fields, even in regions of flat bands,\cite{saito-mebocn1994}
resulting in quantum-dot-like physics.\cite{moldoveanu-ceottpoqdiasmf2001}

\subparagraph*{Relation to 2D periodic structures.}

It is important to note some similarities, but also some fundamental
differences between the butterfly plots of quasi-1D structures and those in the
original work by Hofstadter\cite{hofstadter-elawfobeiraimf1976} and later
generalizations\cite{rammal-llsobeiahl1985,kreft-moth1996} which handled 2D
periodic structures. Starting out from an analogous Hamiltonian and also using
the Peierls substitution to implement the magnetic field, the most striking
difference is that, for a 2D periodic structure, it is not possible in general
to choose a gauge in such a way that the resulting Hamiltonian has the same
translational symmetry as the underlying system. For rational values of the
magnetic flux per unit cell, one can still find a larger effective unit cell
but, for irrational values, this is not possible at all, which ultimately leads to
the fractal structure of the energy spectrum found by Hofstadter, similar to
that displayed in the lower panel of Fig.~\ref{fig:diameter}. In contrast, the
quasi-1D structure of CNTs results in a fixed number of bands,
leading to a \textit{pseudofractal} spectrum, with the
recursion of self-similarity limited by the transverse length-cutoff of the system.

\subparagraph*{Graphene ribbons.}

Since the recent experimental success in isolating single sheets of
graphene,\cite{novoselov-tgomdfig2005, zhang-eootqheabpig2005}
the exotic Dirac-like electronic structure has become the focus of
several studies. Epitaxially grown graphene has been used to laterally
confine electrons and determine coherence lengths studying
weak-localization effects in magnetotransport measurements.\cite{berger-ecacipeg2006}
For understanding the relation between the butterfly of a 2D graphene
sheet and these quasi-1D carbon nanotubes, it is instructive to take a look
at graphene ribbons as an intermediate step.
An $(n,m)$ graphene ribbon is simply a planar ``unrolled'' $(n,m)$
CNT, periodic in one dimension and finite in the other. As in the original Hofstadter
butterfly,\cite{hofstadter-elawfobeiraimf1976} the ribbon butterfly
plots are periodic as a function of the perpendicular magnetic field due to the equal
flux piercing any hexagonal plaquette forming the honeycomb lattice (see Fig.~\ref{fig:grapheneribbon}).
As the ribbon width increases the
butterfly plots tend to the Hofstadter butterfly of a graphene layer
as visible in the bottom panel of Fig.~\ref{fig:diameter}.

\subparagraph*{Cusp catastrophes.}

\begin{figure}
\includegraphics[width=\columnwidth]{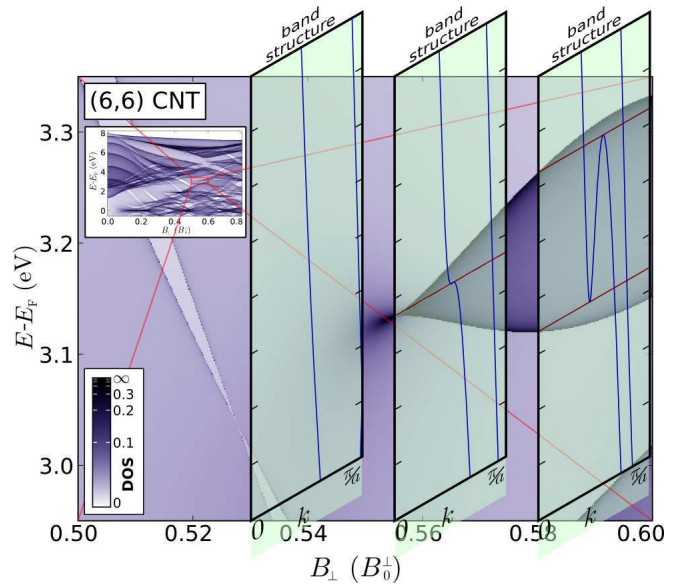}
\caption{\label{fig:cusp}(Color online) 
Cusp catastrophes are ubiquitous in butterfly plots. The band structure at magnetic fields below,
at, and above the critical magnetic field shows the smooth transition from a strictly monotone band
into a third-order parabolic band with changing magnetic field.
}
\end{figure}

One striking detail ubiquitous in butterfly plots are
the cusp catastrophes appearing at specific positions of 
energy and magnetic fields (see Fig.~\ref{fig:cusp}). These are points where,
with changing magnetic field, some band is continuously deformed from a
strictly monotonic curve into a band with two adjacent zero-group-velocity points. At the exact
point where this mathematical catastrophe happens, both the first
and the second derivatives of $E (k)$ are zero. A wave packet of this energy
and momentum will have both its velocity and its spreading suppressed leading
to a special kind of localization not unlike that of Landau levels.

\section{Single-wall carbon nanotubes}

For SWCNTs we consider only the radial
$p$ orbitals---forming the most electronically relevant $\pi$ bands---and only
interactions between nearest neighbors, setting $\gamma_{i j}^0 = \gamma_0 =
2.66 \operatorname{eV}$. This has been shown to be an excellent
approximation in explaining electronic structural and transport properties of SWCNTs.\cite{saito-ppocn1998,reich-cnbcapp2004,thune-qticn2005} The
on-site energy $\varepsilon_i=\varepsilon_0$ is constant for all atoms and defines the Fermi energy
$E_{\mathrm{F}}=\varepsilon_0$ of a neutral CNT.
Ignoring an offset in the energy, we can simply choose $\varepsilon_0 = 0$.
Zeeman splitting could also easily be included in this calculation as
$\varepsilon_0={\pm} g \mu_\mathrm{B} B/2$ and would result visually in an overlay of two
butterfly plots sheared against each other linearly with growing magnetic fields.
The intensity of this effect at the critical plaquette field scale is
$g \mu_\mathrm{B} B_{\text{plaquette}} = 9.1 \operatorname{eV}$.

\subparagraph*{The special case of parallel magnetic fields: Shortcut via the zone-folding method.}

As an alternative to calculating the electronic bands of a SWCNT via the
procedure described above, one could calculate the spectrum of graphene
and then apply periodic boundary conditions in the angular
direction of the CNT (\textit{zone folding}). For magnetic
fields parallel to the tube axis, this method is still applicable:
the phase gathered by an electron moving on a closed loop around the
tube circumference can simply be included in the boundary
conditions. This results in a shift of the allowed discretized
quasimomenta in the reciprocal space. For perpendicular magnetic
fields, however, this method breaks down and one has to consider the
full geometry of the CNT.

\subsection{Structural properties}

\begin{figure}
\includegraphics[width=\columnwidth]{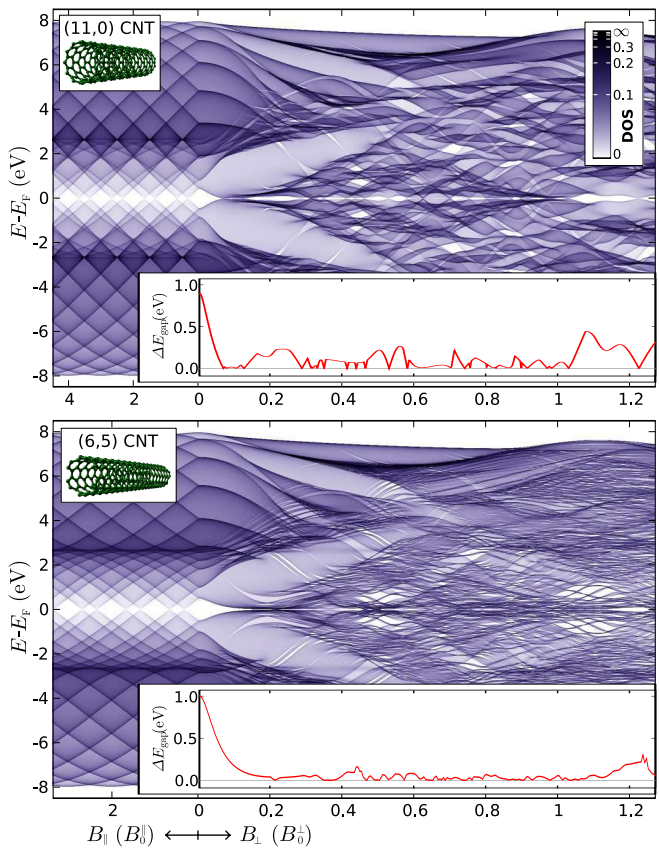}
\caption{\label{fig:chirality}(Color online) Two semiconducting SWCNTs of similar
diameter as the (6,6) CNT in Fig.~\ref{fig:firstplot}. The band gap oscillates
irregularly with increasing perpendicular
field. The large unit cell of the the chiral (6,5) tube leads to bands with
low dispersion as soon as the rotational symmetry is broken by the
perpendicular magnetic field.
}
\end{figure}

\subparagraph*{Chirality dependence.}

Several features can be found when comparing the magnetic spectra of
tubes with different chiralities though similar diameter (see
Figs.~\ref{fig:firstplot} and \ref{fig:chirality}).
(i)
The behavior of the gap around the charge neutrality point
$E=E_{\mathrm{F}}$ is very helicity dependent: a parallel
magnetic field always opens and closes the gap periodically as a
consequence of the integer number of fluxes per nanotube
cross section. This phenomenon is independent of whether the tube is metallic
or semiconducting at $B = 0$. In contrast, for perpendicular fields
there are distinctions: Armchair CNTs stay strictly metallic
for any perpendicular field, as can be understood from supersymmetry
arguments.\cite{lee-sicniatmf2003} On the other hand, the gaps of
the two semiconducting CNTs in Fig.~\ref{fig:chirality} do
open and close in an aperiodic, though oscillatory, pattern. The gap
closes to zero in single points of specific values of $B_{\perp}$
and opens again. Closer observations of a larger set of CNTs reveal
that this also happens for semimetallic tubes like the $(3 n,
0)$ zigzag CNTs. The precise opening and closing pattern carries an
intrinsic complexity; its statistical behavior, however, seems to
depend mostly on the number of atoms in the unit cell.
(ii)
Another general effect of the large unit cell in
the $(6, 5)$ CNT is that in this chiral tube with its large number of
plaquettes at different angles towards the field, the high symmetry of
the original system is broken down very efficiently by the magnetic
field, resulting in a larger number of bands of very low dispersion.
The magnetic field effectively
localizes the electrons in nonpropagating Landau-like states.

\begin{figure}
\includegraphics[width=\columnwidth]{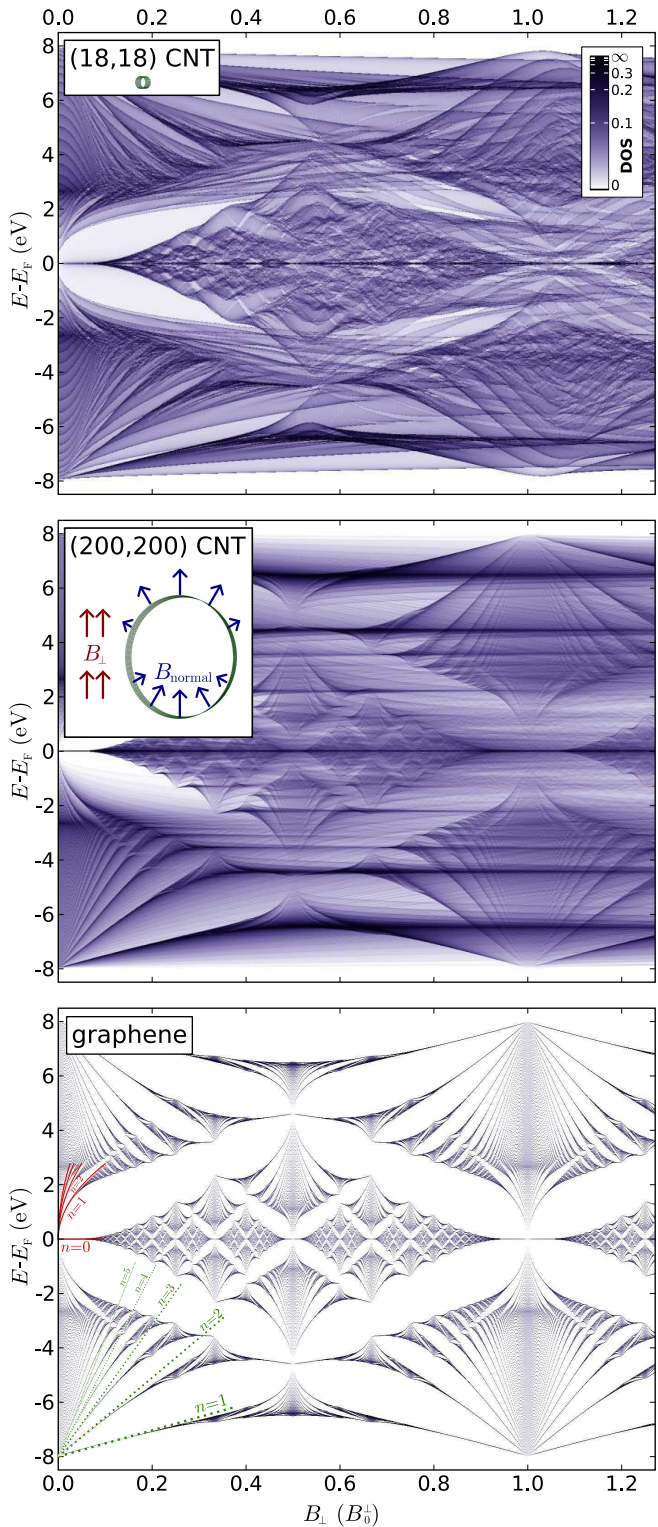}
\caption{\label{fig:diameter}(Color online) Comparing
different diameters: the large (200,200) CNT bears strong resemblance to the
Hofstadter butterfly of graphene (Ref. \onlinecite{kreft-moth1996}) combined with the
curvature effects (details in text).
The straight lines at the lower left corner of the graphene butterfly (bottom panel) indicate
the Landau states obtained from an effective mass continuum theory [see Eq.~(\ref{eqn:landau-levels})].
The parabolic lines near $E_{\mathrm{F}}$ in the same plot indicate the relativistic Landau levels
obtained from the Dirac-like dispersion of graphene [see Eq.~(\ref{eqn:relativistic-landau-levels})].
}
\end{figure}

\subparagraph*{Diameter dependence.}

Figure~\ref{fig:diameter} shows the evolution toward the
graphene Hofstadter butterfly of the magnetic spectrum of armchair
CNTs as a function of their diameter. The (200,200) SWCNT has a
diameter of 27~nm, comparable to the external shell of a typical MWCNT, and is thus
of great interest. The visual effect---resembling
watercolors ``flowing'' toward the right hand side---can also be
understood by a simple picture. For large enough diameters, the
CNT consists of regions of nearly flat graphene, each at a different
angle toward the magnetic field, thereby experiencing a different
normal component of the magnetic field, as visible in the inset of the middle panel of Fig.~\ref{fig:diameter}.
Since the DOS is an average over the
LDOSs at the different unit cell atoms, one ends up with
a sum of different graphene Hofstadter butterflies, stretched to different
effective fields, the stretch being minimal where the magnetic field is normal to the
tube wall and maximal where it is tangential. Overlaying these differently
stretched graphene butterflies results in the ``flowing'' appearance
of the butterfly of large diameter tubes.

At the lower and upper energy edges, one can clearly see the
emergence of linear Landau levels and the characteristic fractal
structure of the graphene butterfly is unmistakenly visible at the
same scale of the magnetic field. In fact near the top and
bottom of the graphene $\pi$~energy band
(of width $2 W=6 \gamma=16~\mathrm{eV}$), electrons have an effective
mass of $m^{*} = 2\hbar^2/3\gamma d_\mathrm{CC}^2 \approx 0.95 m_\mathrm{e}$, leading to a cyclotron frequency
$\omega \lrb B_{\perp} \rrb=e B_{\perp} /m^{*}$, so one could write
\begin{eqnarray}
E&=&\mp W \pm \hbar \omega (B)\lrb n + \frac 1 2 \rrb\label{eqn:landau-levels}
\end{eqnarray}
with $n=0,1,2,\dots$, which fits nicely with the numerical results (as indicated by the straight lines
in the bottom panel of Fig.~\ref{fig:diameter}).

Around the Fermi energy, the Dirac-like dispersion of graphene leads to the so-called
relativistic Landau levels,\cite{zheng-hcoatgs2002,haldane-mfaqhewllcrota1988} following
\begin{eqnarray}
E&=&\pm v_{\mathrm{F}} \sqrt{2 n \hbar e B_\perp}\label{eqn:relativistic-landau-levels}
\end{eqnarray}
with $n=0,1,2,\ldots$ and the Fermi velocity $v_{\mathrm{F}} = 3 \gamma d_{\mathrm{CC}} / 2\hbar$.
These levels, which can be clearly observed in the Hofstadter butterfly of graphene
(left edge in the bottom panel of Fig.~\ref{fig:diameter}), are also responsible for the recently
observed anomalous quantum Hall effect of graphene.\cite{novoselov-tgomdfig2005, zhang-eootqheabpig2005}

\begin{figure*}
\includegraphics[width=\textwidth]{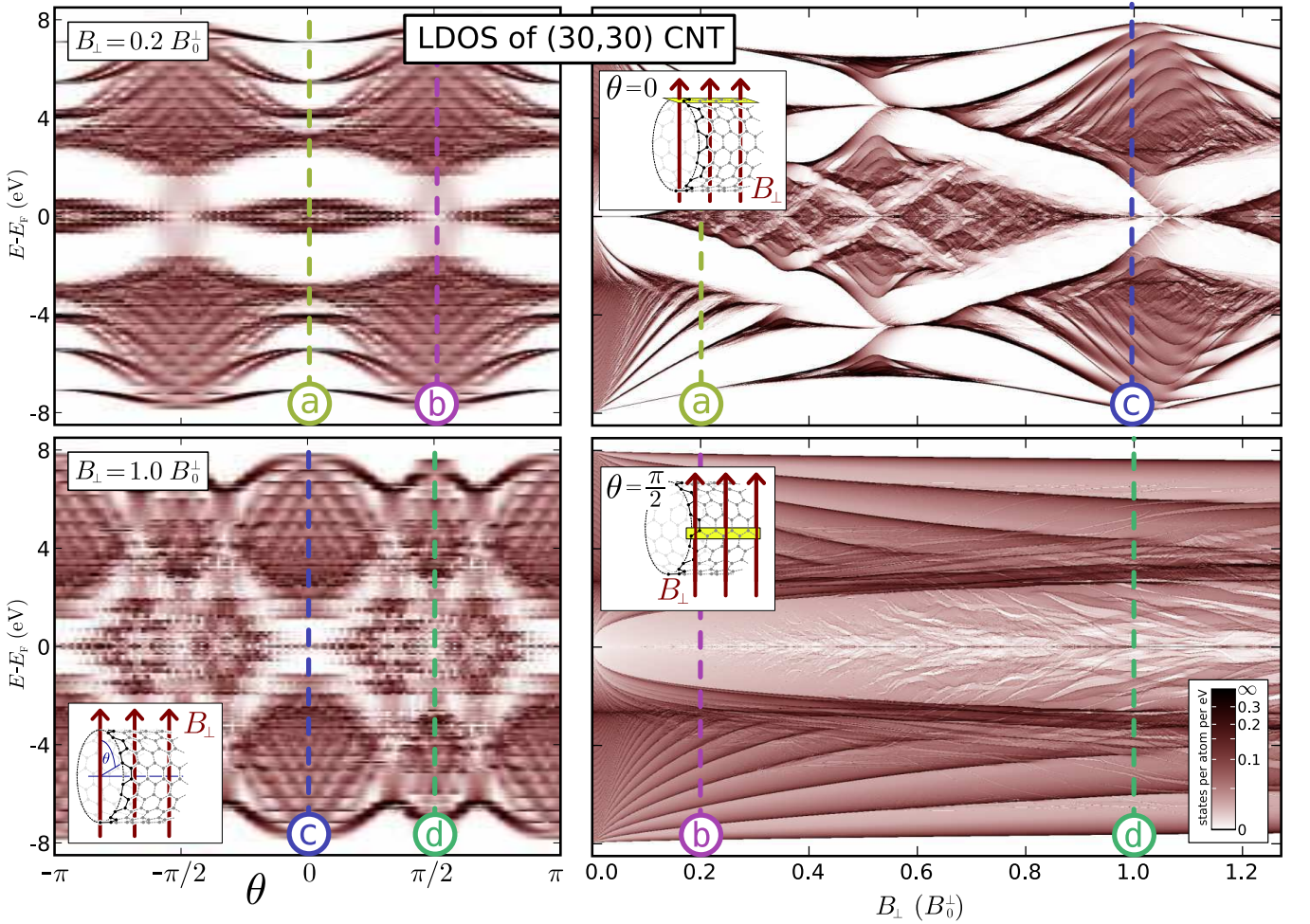}
\caption{\label{fig:ldos}(Color online) Decomposition
of the density of states into the contributions of particular atoms (identified
by their angle $\Phi$ toward the magnetic field direction).
A plaquette at angle $\theta$ captures a flux of $B_{\perp} A_{\text{plaquette}}
\cos \theta$.
The region at $\Phi = 0$ experiences a perpendicular field piercing the tube
wall, very much as in the plain graphene sheet (Fig.~\ref{fig:diameter}).
The regions at $\Phi = \pi/2$ experience a field tangential
to the tube wall, leading to a much smaller flux per plaquette, resulting a the
stretched impression of the butterfly. The DOS butterfly over the
whole CNT unit cell is an overlay of these and many intermediate
pictures.
In the angle-resolved plot for $B = 0.2 ~ B_0^\perp$
one can see a smooth transition between a region
with Landau levels and a region with normal band dispersion. For the stronger
field, the systems goes through two oscillations along the angle.
}
\end{figure*}

\subparagraph*{Snake states.}

The view of the total DOS of a large SWCNT as the sum of different contributions from the regions at various angles around the tube circumference can be confirmed by taking a look at the LDOS at individual atom positions. As can be seen in Fig.~\ref{fig:ldos}, the LDOS at $\theta = 0$, where the magnetic field pierces the wall perpendicularly, resembles very much the butterfly of the planar graphene sheet. The electrons here show very low dispersion, similar to Landau levels. At $\theta = \pi/2$, on the other hand, the magnetic field is tangential to the CNT wall and therefore has far less effect on the electron dispersion. An understanding of the electronic states in these regions can be gained by considering classical electrons confined to the surface of a cylinder: As the effective magnetic field (the projection of the field onto the tube normal) changes sign at $\theta = \pi/2$, the curvature of an electron trajectory will also switch orientation each time the electron crosses this ``equator'' line, leading to a snakelike
movement of the electron.\cite{lee-sicniatmf2003,mller-eoanmfoategitbr1992}

\subsection{Analytical model}

\begin{figure}[t]
\includegraphics[width=\columnwidth]{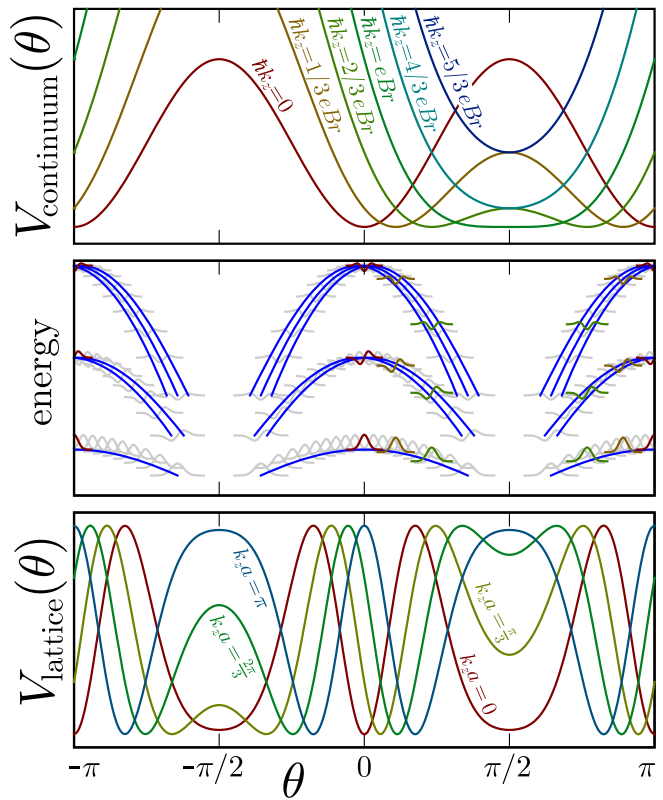}
\caption{\label{fig:cylinder}(Color online) Analytical solution of the continuum model. Top panel:
the $k_z$-dependent effective potential of free electrons confined to a continuum cylinder in a
magnetic field perpendicular to the tube. Middle panel: first three eigenstates of the harmonic
approximation to the above potential for various $k_z$ (highlighted wave functions correspond to the
selected potentials in the first panel). Each
wave function is shifted to the corresponding energy. Superimposed are the lines followed by
the extrema of the wave functions. The same pattern can be found in the top left panel of 
Fig.~\ref{fig:ldos}, where the maxima of the DOS show the maxima of the various energy eigenstates.
Bottom panel: The $k_z$-dependent effective potential of a discretized tube showing a large number
of minima. States located in narrow minima have higher energy, so the low-energy spectrum is
mainly determined by the widest potential minima.
}
\end{figure}

In order to shed some light of intuition on our results, we may consider the
physics of a structureless hollow cylinder, a tubule, in a perpendicular
magnetic field. (Similar systems in parallel magnetic fields have been studied before.{\cite{margulis-etoacswol2005,zavyalov-eohmfotcoaqcuslc2005}})
This system bears some similarity to a ``Hall bar'', with the crucial
difference that it does not have borders that could carry edge states.
Instead, it has two flanks where the magnetic field is tangential to the tube
and therefore the radial component of the magnetic field---which is the
effective field experienced by electrons confined to the cylinder surface---vanishes.
To understand where charges do accumulate, we consider this system
in cylindrical coordinates $(\theta, z)$ at fixed radius $r$. By
Eq.~(\ref{gauge-field}), a perpendicular magnetic field leads to a gauge field
\[ \mymathbf{A} \left( \theta \right) = B_{\perp} r \sin \theta ~ \mymathbf{e}_z \]
in cylindrical coordinates. With this, the Hamiltonian of an electron
restricted to the tube surface becomes
\[ \mathcal{H} = \frac{1}{2 mr^2} p_{\theta}^2 + \frac{1}{2 m} \left( p_z -
   eB_{\perp} r \sin \theta \right)^2 \]
which can be viewed as that of an electron in 2D with periodic boundaries in a
nonuniform magnetic
field.{\cite{handrich-qmmdvaasm2005,krakovsky-ebsiapmf1996}} A similar 
system---a 2D strip ranging over $[- L / 2, L / 2]$ in the $y$ direction and
infinite in the $z$ direction, placed in a linearly varying magnetic field
$\mymathbf{B} = B_0 y \mymathbf{e}_x$---was first studied in 1992 by
M\"uller,{\cite{mller-eoanmfoategitbr1992}} who identified two new classes
of states: one at finite magnetic field propagating perpendicularly to the field
gradient direction with looping trajectory and low velocity, the other around
the line $B = 0$, propagating in the opposite direction at higher velocity
with a snakelike trajectory. To solve our system, we can exploit the
$z$ invariance and do an ansatz for the wave function: $\Psi \left( \theta, z
\right) = \psi_{k_z} \left( \theta \right) \mathrm{e}^{\mathrm{i} k_z z}$.
Our problem reduces to that of a particle in one dimension with a $k_z$ dependent
potential:
\begin{eqnarray}
  \mathcal{H}_{k_z} & = & \frac{1}{2 mr^2} p_{\theta}^2 + V_{k_z} \left(
  \theta \right) \nonumber\\
  V_{k_z} \left( \theta \right) & = & \frac{1}{2 m} \left( \hbar k_z -
  eB_{\perp} r \sin \theta \right)^2. \label{potential-of-continuum-model}
\end{eqnarray}
For $\left \vert \hbar k_z \right \vert < \left \vert e B_{\perp} r \right \vert$, this potential
has two minima at $\theta_{\min} = \pi / 2 \pm \arccos \left( \hbar k_z /
e B_{\perp} r \right)$. A harmonic approximation at either of these minima
yields the approximate Hamiltonian
\begin{eqnarray*}
  \mathcal{H}_{k_z} & = & \frac{p_{\theta}^2}{2 mr^2} + \frac{1}{2 m} \left[ \left(
  eB_{\perp} r \right)^2 - \left( \hbar k_z \right)^2 \right] \left( \theta -
  \theta_{\min} \right)^2
\end{eqnarray*}
with the spectrum
\begin{eqnarray*}
  E_n \left( k_z \right) & = & \frac{\hbar}{m r} \sqrt{\left(
  eB_{\perp} r \right)^2 - \left( \hbar k_z \right)^2} \left( n + 1 / 2
  \right).
\end{eqnarray*}
From this dispersion relation, we can directly retrieve the group velocity
\begin{eqnarray}
  v_n \left( k_z \right) & = & - \frac{\hbar}{m r} \frac{\left( n + 1 /
  2 \right) \hbar k_z}{\sqrt{\left( eB_{\perp} r \right)^2 - \left( \hbar k_z
  \right)^2}}. \label{group-velocity}
\end{eqnarray}
The wave functions in the harmonic potential are located around the minima
$\theta_{\min}$, so for low energies we can say in reverse that at each angle $\theta$ we
find predominantly electrons with the longitudinal wave vector $k_z \left( \theta \right) =
\left(e B_{\perp} r / \hbar \right) \sin \theta$. Placing this into
Eq.~(\ref{group-velocity}), we can retrieve an expression for the velocity of
electrons moving at certain angles:
\begin{eqnarray*}
  v_n \left( \theta \right) & = & - \frac{\hbar}{m r} \left( n + 1 / 2
  \right) \tan \theta.
\end{eqnarray*}
Now, the divergency at $\theta = \pm \left( \pi / 2 \right)$ originates from the fact
that the harmonic approximation breaks down when the two minima of $V_{k_z}
\left( \theta \right)$ meet at this angle. Apart from this, however, one can
see clearly the angular separation of electrons moving in both directions and
the localization in Landau-like states at $\theta = 0$ and $\theta = \pi$,
where the magnetic field pierces the tube wall normally (see Fig.~\ref{fig:cylinder}).

Such a continuum model can only be expected to hold for the CNTs
at low magnetic fields with $B_{\perp} A_{\operatorname{plaquette}} \ll \Phi_0$. As
it turns out, some of the effects visible at higher fields can be understood
qualitatively by studying a model of intermediate complexity: a square lattice 
cylindrical tube of lattice constant $a$. Coming from the
continuum model and following Ref.~{\onlinecite{yi-mocn2004}}, we can replace
the continuous coordinates by integer indices: $\left. (am, an \right) :=
(r \theta, z)$.
Using a tight-binding model with on-site energy $\varepsilon_0$ and hopping
parameter $\gamma_0$, the Hamiltonian acts on a wave function in the following
way:
\begin{eqnarray*}
  \mathcal{H} \Psi_{\left( m, n \right)} & = & \varepsilon_0 \Psi_{\left( m, n
  \right)}\\
  & - & \gamma_0 \left( \Psi_{\left( m - 1, n \right)} + \Psi_{\left( m + 1, n
  \right)} \right)\\
  & - & \gamma_0 \left( \operatorname{e}^{- \mathrm{i} \varphi \left( m \right)}
\Psi_{\left(
  m, n - 1 \right)} + \operatorname{e}^{\mathrm{i} \varphi \left( m \right)} \Psi_{\left( m,
  n + 1 \right)} \right)
\end{eqnarray*}
where the phase factor $\varphi \left( m \right) = \left( 2 \pi
a r B_{\perp} / \Phi_0 \right) \sin \left( ma / r \right)$ originates from the Peierls
substitution Eq.~(\ref{peierls}). As in the continuum, the invariance in the
$z$ direction can be exploited, now using a Bloch ansatz due to the
discreteness of the system:
\begin{eqnarray*}
  \Psi_{(m, n)} & = & \operatorname{e}^{\mathrm{i} k_z an} \psi_m~.
\end{eqnarray*}
This leads to a finite Hamiltonian for any fixed $k_z \in \left[ - \pi / a, \pi /
a \right)$:
\begin{eqnarray*}
  \mathcal{H}_{k_z} \psi_m & = & - \gamma_0 \left( \psi_{m - 1} + \psi_{m + 1}
\right) + V_{k_z} \left( m \right) \psi_m,
\\
  V_{k_z} \left( m \right) & = &
    \varepsilon_0 - 2 \gamma_0 \cos \left( k_z a - 2 \pi \frac{r a
B_{\perp}}{\Phi_0} \sin \frac{m a}{r} \right).
\end{eqnarray*}
The most significant difference to the effective potential of the continuum
model Eq.~(\ref{potential-of-continuum-model}) is the replacement of the square law
by a cosine one. This has the effect that the potential does not grow indefinitely
for large magnetic fields, but instead oscillates, forming several minima at
various angles $\theta$, as seen in Fig.~\ref{fig:cylinder}. In combination with the discretization of the angle,
this potential leads to the formation of a complex pattern in the angular dependence
of the density of states, as it can be observed is the LDOS at high magnetic
fields also displayed in Fig.~\ref{fig:ldos}.

To capture more details in a model, an appropriate step would be the implementation
of the correct dispersion at the Fermi energy: The characteristic cones at the Fermi points
of graphene can be approximated by a Dirac-like Hamiltonian. For a detailed study of the magnetic
spectrum of Dirac-electrons on a cylindrical surface, see [\onlinecite{lee-sicniatmf2003,perfetto-qheicn2006}].

\subsection{Experimentally accessible perpendicular magnetic fields}

\begin{figure}
\includegraphics[width=\columnwidth]{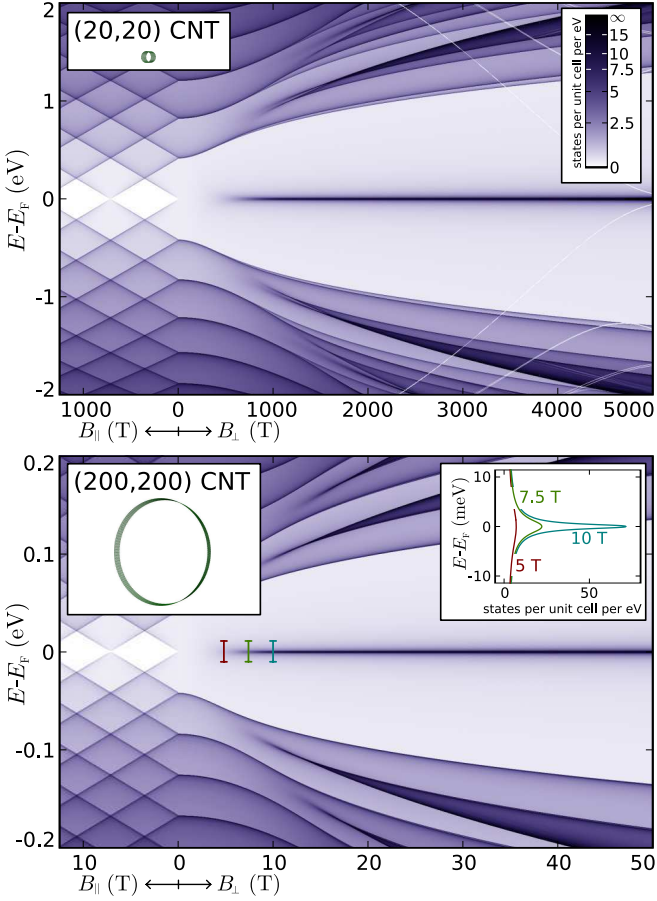}
\caption{\label{fig:lowfield}(Color online) Zoom into the butterfly of two different
armchair SWCNTs. The scales, including the color scale,
are chosen according to the scaling law given in the text to produce
comparable representations of the data.
The tubes correspond to diameters of 2.7~nm (top) and 27~nm (bottom).
The inset in the bottom panel illustrates the shape of the peak at the Fermi level.
The white lines crossing the plot in the upper panel are caused by small avoided
crossings in the band structure.}
\end{figure}

In recent experiments, optical transitions in CNTs were studied in magnetic
fields up to 75~T.\cite{zaric-eicnwbts2006}
New experiments are in preparation to go up to 200~T and even
2--3~kT.\cite{kono-2005}

In perpendicular fields of this magnitude, as displayed in
Fig.~\ref{fig:lowfield}, the first onset of the band structure distortions can
be seen clearly in large CNTs, comparable with the outer shell of
typical MWCNTs, measuring up to tens of nm in diameter.

In particular the plots show very clearly the rapidly changing
van Hove singularities, resembling those of Fig.~\ref{fig:firstplot}
for a (6,6)~CNT, but at much lower magnetic field scale. Moreover, the
$E=E_\mathrm{F}$ graphene state, which is due to the peculiar distortion of the
Dirac-like linear dispersion into a strongly nonlinear
one,\cite{roche-mocnfmtmf2001} emerges at lower fields with increasing
diameters.

Most notable is the scaling law that can be found in the butterfly plot of
large tubes at low fields near the Fermi energy: For two different armchair
CNTs with the chiral vectors $(m,m)$ and $(m',m')$ it can be expressed as
\begin{eqnarray*}
\rho_{\operatorname{DOS}_{(m,m)}}\left(E,\mymathbf{B} \right) &=& \frac{m'}{m}
\rho_{\operatorname{DOS}_{(m',m')}}\left(\frac{m}{m'}E,\frac{m^2}{m'^2} \mymathbf{B}\right).
\end{eqnarray*}

This scaling is followed approximately already for
small CNTs and becomes very precise for large diameters, converging toward a
DOS that is reproducible from a model of Dirac electrons on a continuum
cylinder.\cite{lee-sicniatmf2003} The peak at the Fermi energy also follows this scaling law. Within the
region of scaling, the maximum of
the peak at $E=E_\mathrm{F}$ grows exponentially with the magnetic field while its
integral grows linearly.

It is important to note that the scaling is not an effect of the curvature, but
of the discretization of the transversal momentum, since it can be observed in
graphene ribbons as well.

\section{Double-wall carbon nanotubes}

While SWCNTs and MWCNTs have been studied
intensely over the past 15 years, it has only recently become possible to
produce DWCNTs of high purity and quality,\cite{hutchisona-dcnfbahadm2001, sugai-nsohdcnbhpad2003} fueling the interest in
details about the interwall interaction. Previous studies have shown an interesting interplay between magnetic fields
parallel to the DWCNT axis and the interwall interaction near the Fermi energy.\cite{ho-bsodcn2006}
A minimal Hamiltonian of a DWCNT can be set up as
\begin{eqnarray*}
\mathcal{H}&=&\sum_{\left<i,j\right>} \gamma_{i j}\left(\mymathbf{B}\right) c_i^\dagger c_j^{\phantom{\dagger}}
+ \sum_{\left<\left<i,j\right>\right>} \tilde{\gamma}_{i j}\left(\mymathbf{B}\right) c_i^\dagger c_j^{\phantom{\dagger}}
\end{eqnarray*}
by defining the intrawall interaction as described
for SWCNTs. For the interwall interaction, we can fix the hopping
coefficients as
\begin{eqnarray*}
\tilde{\gamma}_{i j}\left(\mymathbf{B}\right) &=&
\beta \cos \vartheta_{i j} \exp \left( \frac{d_{i j} -
a}{\delta} \right) \\ & & \times \exp \left[  \mathrm{i}
\frac{2 \pi}{\Phi_0} \mymathbf{d}_{ij} \cdot
\mymathbf{A}_{\mymathbf{B}} \left( \frac{\mymathbf{r}_j + \mymathbf{r}_i}{2} \right)
\right],
\end{eqnarray*}
where $\beta = \gamma_0 / 8$, $a = 3.34$~{\AA}, $\delta = 0.45$~{\AA}, and
$\vartheta_{i j}$ and $d_{i j}$ stand for the angle and the absolute
distance between the two $\pi$ orbitals $\left<\left<i,j\right>\right>$ centered at
positions $\mymathbf{r}_i$ and $\mymathbf{r}_i$ belonging to two different shells.\cite{lambin-esocct1994,roche-cmamimcn2001}

\begin{figure}
\includegraphics[width=\columnwidth]{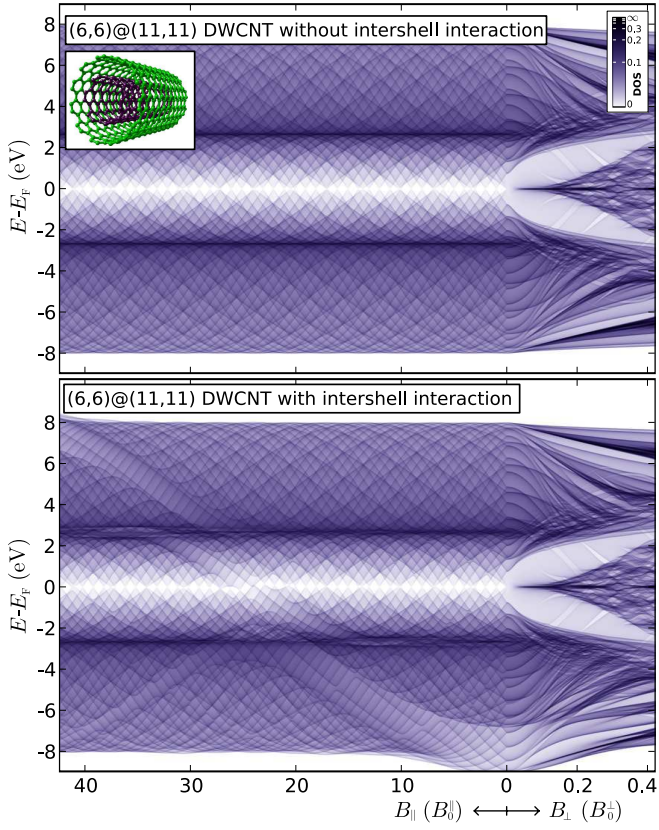}
\caption{\label{fig:dwcnt}(Color online) Butterfly plot of a (6,6)@(11,11) double-wall CNT.
In the upper panel, the interwall interaction is switched off, resulting
in an overlay of the butterflies of two independent SWCNTs. In the lower panel,
the interwall interaction gives rise to a number of new features (see text for
details).}
\end{figure}

\begin{figure}
\includegraphics[width=\columnwidth]{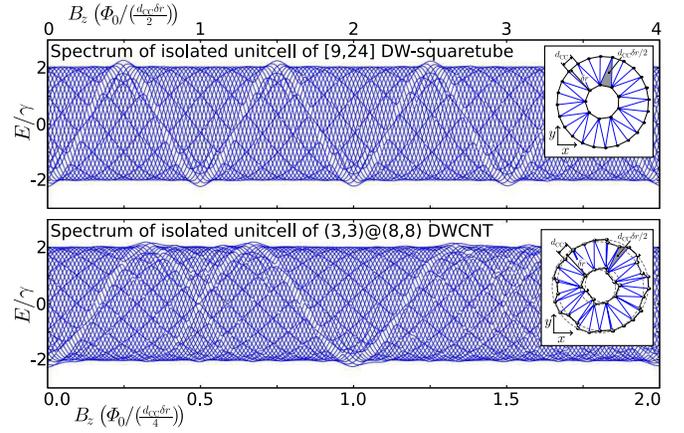}
\caption{\label{fig:doublering}(Color online) Upper panel: Spectrum of a system of two concentric atomic rings.
Atom spacing and coupling inside each ring are taken from graphene. The
distance $\delta r$ between the rings as well as the the parametrization of the
coupling between the rings follow those given in the text for DWCNTs. The sketch displays
the prevalent links between the shells. Even though the geometry is irregular, the area of
circular paths is very near to integer multiples of $d_\mathrm{CC} \delta r/2$,
leading to a clear periodicity of the modulation in the spectrum.
Lower panel: An isolated unit cell of a DWCNT with the same radii as the planar double ring of the upper panel.
This system has smallest closed loops at an angle against the magnetic field, resulting in
an effective smallest area of $d_\mathrm{CC} \delta r/4$ and leading to a doubling of the period. Furthermore, the
system has two atoms in the rotational periodic cell, leading to two interlaced modulations.}
\end{figure}

As a representative example, the butterfly of a (6,6)@(11,11) DWCNT is displayed in Fig.~\ref{fig:dwcnt}.
For the parallel magnetic field, the interwall interaction gives rise to features at
two levels: looking at individual bands, one can observe van Hove singularities crossing
and sometimes avoid a crossing. The complex rules whether a crossing is avoided are not
obvious from studying the butterfly plot only, but can be deduced by looking at the
band structure and taking into account the various symmetries of the system.
At larger scale in the butterfly plot, one finds a
modulation of the pattern crossing from $E = E_\mathrm{F} - 8 \operatorname{eV}$ at $B_{\parallel}=0$ to
$E = E_\mathrm{F} + 8 \operatorname{eV}$ at $B_{\parallel} \approx 45~B_0^{\parallel}$.

To understand this phenomenon, we have studied a single unit cell in a magnetic field perpendicular to the plane of the
resulting concentric ring (Fig.~\ref{fig:doublering}). The spectrum shows a periodic
behavior of the modulation with a period
$B_0^{\mathrm{interwall}} = \Phi_0/\left (d_\mathrm{CC} \delta r_\mathrm{interwall} / 4\right )$
relating to the area of the minimal circular path between both walls. This period and the shape of
the modulation are independent of the diameter of the DWCNT.
A physical explanation for this modulation is as follows. Like the interaction in a two-atomic
molecule, splitting two atomic orbitals into a bonding and an antibonding molecular orbital,
the interwall interaction may also hybridize SWCNT states of the same energy into bonding and 
antibonding DWCNT states. However, the system has an approximate rotational symmetry, so the
interwall interaction may only hybridize states of the same angular momentum.

At zero magnetic
field, the angular momentum of the states at the bottom of the spectrum is zero in both shells.
This allows hybridization, causing a split in the hybrid spectrum. At the upper end of
the spectrum, the angular momentum does not match, prohibiting a hybridization. This is the
cause for the strong electron-hole asymmetry visible in DWCNT butterfly plots.

By switching on a parallel magnetic field, the effective angular momentum is shifted by the
Aharonov-Bohm phase gathered on a circular path around the tube. This shift depends on the cross section
of the path, so it is different for the two shells. Therefore, the energy at which both angular
momenta match depends on the magnetic field, causing the splitting region to travel over the energy range,
which leads to the visible modulation in the parallel field butterfly plots.

For a simplified model---a double-wall square lattice tube---the modulation does follow a single cosine-shaped curve,
as shown in the upper panel of Fig.~\ref{fig:doublering}.
In comparison, the DWCNT shows an additional complexity: the underlying honeycomb lattice of graphene has a
unit cell containing two atoms, resulting in two intertwined cosine curves, the second just becoming visible at the edge of
Fig.~\ref{fig:dwcnt}.

For fields perpendicular to the axis of a DWCNT, the only large-scale effect caused by the interwall
interaction observable in the
butterfly plot is the hybridization-induced splitting already described for zero field.
With growing field, this effect disappears, and the
plot shows no remarkable global patterns.

\section{Conclusions}

The magnetic spectrum of two-dimensional
infinite lattice electrons gives rise to the
well-known Hofstadter butterfly. In this paper,
we have shown that quasi-one-dimensional lattice
electrons exhibit a spectrum which does resemble
the fractal structure of the Hofstadter
butterfly but with a finite cutoff due to the
transversal confinement. We have calculated
such pseu\-do\-frac\-tals for carbon nanotubes,
a material at the focus of many nanoelectronic
studies also in relation to the presence of
external magnetic fields. We have calculated
the density of states (butterfly plots) of
several single wall carbon nanotubes and we could show
(i) the strong dependence of the magnetic spectrum
on the underlying  chiral indices;
(ii) the emergence of the graphene Hofstadter butterfly
at increasing nanotube diameter.
In particular, perpendicular fields induce an
aperiodic and pseudofractal magnetic
spectrum. Periodic structures have been
obtained for graphene ribbons, demonstrating that
the aperiodicity of the perpendicular field
butterfly plots is due to the
incommensurability of the magnetic flux captured by
elementary (hexagon) plaquettes of a CNT oriented
at different angles towards the external field.

By studying the angle-resolved electronic structure
of a SWCNT one can observe the emergence
of snake states already predicted for
nonuniform magnetic fields in a Hall bar.\cite{mller-eoanmfoategitbr1992}
In our case, we have been able to devise an
analytical model for the states at the top and the bottom
of the energy spectrum by means of an effective mass
approximation.  In this latter case a continuum
theory can capture the striping of the wave
function along the region of the tube with zero
normal field. Inversely, near the Fermi level,
one cannot bypass the Dirac neutrino nature of the
electronic states. We have interpreted the wave function striping
by writing a Harper equation\cite{harper-sbmoceiaumf1955}
for square lattice electrons with a cylindrical geometry.

While the effects of parallel fields are of
comparably simple nature in SWCNTs (being an
expression of the Aharonov-Bohm oscillations
due to a rigid shift of the graphene band
structure sampled via the zone-folding method),
this is not the case for DWCNTs. The
electron-hole symmetry of $\pi$ bands in
SWCNTs is broken once two shells are put in
interaction. The resulting hybridization of
inner and outer states could be clearly
understood by means of two interacting
Aharonov-Bohm rings.

Experimentally relevant effects have
been calculated for SWCNTs of diameter of
typical external shells in MWCNTs. There
underlying multifractal structure like that of
Hofstadter can be observed already at a few tens
of tesla, and an outstanding scaling law for the DOS at low
magnetic fields near the Fermi energy has been given.
The latter applies also to graphene ribbons and is
intrinsically related to the massless dispersion at the
charge neutrality point.

This study, though systematic, could not include
very interesting issues which also deserve
careful investigation, such as the effects of
disorder on the butterfly plots of SWCNTs.
Disordered SWCNTs can be thought in fact a model for the
external shell of MWCNTs. More atomistically
one could study the influence of the interwall
interaction of the structure of large diameter
DWCNTs (also reasonable models for
MWCNTs).\cite{nemec-2006}

\section*{Acknowledgments}

We acknowledge fruitful discussions with J.~Kono, S.~Krompiewski, E.~Heller, U.~R\"o{\ss}ler, and
C.~Strunk. Ulrich R\"o{\ss}ler made us aware of the interesting history
of science involving Douglas Hofstadter, Gregory Wannier, and Gustav Obermeir which took
place in Regensburg during the mid 1970s and eventually led to the
discovery of the Hofstadter butterfly thirty years ago. This work was funded
by the Volkswagen Foundation under grant No.~I/78~340,
by the European Union grant CARDEQ under Contract No.~IST-021285-2
and by the Deutsche Forschungsgemeinschaft within the Collaborative Research Center SFB 689.
Support from the Vielberth Foundation is also gratefully acknowledged.

\appendix

\section{\label{app:histogram}Histogram method}

A histogram method is the simplest method to get the magnetic spectrum of a quasi-1D system.
It is also very efficient if the complete energy range
has to be calculated. Starting from a periodic
Hamiltonian of the form:
\begin{eqnarray}
\mathcal{H} & = & \left( \begin{array}{llllll}
\ddots & \ddots & \ddots & & & \\
& H_1^{\dag} & H_0 & H_1 & & \\
& & H_1^{\dag} & H_0 & H_1 & \\
& & & \ddots & \ddots & \ddots
\end{array} \right), \label{eqn:periodic-hamiltonian}
\end{eqnarray}
one can use Bloch theorem to get an effective Hamiltonian:
\begin{eqnarray*}
H^{\phantom{\dag}}_{\mathrm{\operatorname{eff}}} \left( k \right) & = &
H_0^{\phantom{\dag}} + \mathrm{e}^{\mathrm{i} ka} H_1^{\phantom{\dag}} + \mathrm{e}^{-
\mathrm{i} ka} H_1^{\dag}
\end{eqnarray*}
where $a$ is the length of the unit cell. Numerically scanning
the 1D Brillouin zone $- \pi / a < k \leqslant \pi / a$ with a uniform
distribution, one can now diagonalize the finite matrix
$H_{\mathrm{\operatorname{eff}}} \left( k \right)$ for each value $k$. The resulting
eigenvalues from this diagonalization are counted in a linear histogram over
the full energy range and normalized to the total number of states. Depending
on the resolution of the $k$ sampling, this histogram will become an
arbitrarily good approximation to the density of states. Figures \ref{fig:firstplot}, \ref{fig:chirality},
\ref{fig:diameter} (CNT panels), \ref{fig:grapheneribbon}, and \ref{fig:dwcnt}
were calculated using this method.

The calculation of the data in Fig.~\ref{fig:lowfield} was heavily optimized by using an
adaptive $k$ sampling in combination with a linear interpolation to reduce the number of
diagonalizations in regions of smooth band structure and increase the precision at band edges.

\section{\label{app:greenfcn}Green function method}

Another, more flexible method is that using Green functions: The bulk Green
function $\mathcal{G} \left( E \right)$ of the infinite CNT can be calculated
very efficiently by the following method.\cite{sancho-hcsftcobasgf1985}

The periodic Hamiltonian in Eq.~(\ref{eqn:periodic-hamiltonian}) is used as starting point of a
recursive decimation scheme:
\begin{eqnarray*}
H_0^{(0)}(E) &=& H_0,
\\
H_{01}^{(0)}(E) &=& H_1,
\\
H_{10}^{(0)}(E) &=& H_1^\dag.
\end{eqnarray*}

With each recursion, the length of the effective unit cell is now doubled by decimating out every
second cell:
\begin{eqnarray*}
H_0^{(n+1)}(E) &=& H_0^{(n)}(E) + H_{01}^{(n)} \gamma^{(n)} H_{10}^{(n)} + H_{10}^{(n)} \gamma^{(n)} H_{01}^{(n)},
\\
H_{01}^{(n+1)}(E) &=& H_{01}^{(n)} \gamma^{(n)} H_{01}^{(n)},
\\
H_{10}^{(n+1)}(E) &=& H_{10}^{(n)} \gamma^{(n)} H_{10}^{(n)},
\end{eqnarray*}
where $\gamma^{(n)} = (E + \mathrm{i}\eta - H_0^{(n)})^{-1}$ and $\eta$ is a small positive numerical value, chosen
smaller than the desired energy resolution but large enough to provide fast convergence and numerical stability.

Convergence is reached for
$n \geq n'$ if $\Vert H_{01}^{(n')}(E) \Vert + \Vert H_{01}^{(n')}(E) \Vert < \epsilon$
for some matrix norm $\Vert \cdot \Vert$ and some small cutoff $\epsilon$. We can then retrieve the
bulk Green function from the converged $H_0^{n'}$ as:
\begin{eqnarray*}
\mathcal{G}_\mathrm{bulk} \left( E \right) &\approx& (E + \mathrm{i}\eta - H_0^{(n')})^{-1}
\end{eqnarray*}

With the original Hamiltonian (\ref{eqn:periodic-hamiltonian}) expressed in a $\pi$-orbital
tight-binding basis, the resulting Green function
$\mathcal{G}_\mathrm{bulk}$ is a matrix in the same atomic basis of one unit cell. Therefore, the
local density of states in each atom is directly given by
\begin{eqnarray*}
\rho_{\operatorname{LDOS}_i} \left( E \right) & = & - \frac{1}{\pi} \mathrm{\operatorname{Im}}
\left[ \mathcal{G} \left( E \right)_{i i} \right],
\end{eqnarray*}
summing up to the $\rho_{\operatorname{DOS}} \left( E \right) = \sum_i \rho_{\operatorname{LDOS}_i} \left( E
\right)$. In the same run, the surface Green functions
$\mathcal{G}_{\mathrm{s}}^{\mathrm{L} / \mathrm{R}} \left( E \right)$ can be
used to calculate the transmission through the system using the Fisher-Lee
relation\cite{fisher-rbcatm1981} with a single unit cell selected as
conductor, as shown in Fig.~\ref{fig:explain}.

The Green function method and the histogram method
both give numerical approximations to the same
mathematical quantity, but their numerical errors are very different: while
the former method tends to give fluctuations that show up as grainy structure
in flat areas of the butterfly plot, the latter suffers from sampling problems
around van Hove singularities. Both errors have to be countered with very high
resolution scanning and down sampling of the data. The data presented in the
figures of this article typically took several hours to weeks of computation
time on standard PCs [Intel(R) Pentium(R) 4, 3 GHz].

The work presented here was done using the following
Open Source~(R)\cite{bezroukov-ossdaastoarovr1999} software:
Python as programming language,\cite{rossum-prm2001}
NumPy (Refs. \onlinecite{ascher-np2001} and \onlinecite{oliphant-gtn2006}) and SciPy (Ref. \onlinecite{jones-sosstfp2001}) for numerical computations,
PyTables for data storage and handling,\cite{altet-pug2002}
matplotlib for data visualization,\cite{barrett-mapppp2004}
inkscape for figure preparation,\cite{inkscape2006}
and TeXmacs for authoring.\cite{texmacs2001}




\end{document}